\documentclass{sig-alternate-2013}
\newfont{\mycrnotice}{ptmr8t at 7pt}
\newfont{\myconfname}{ptmri8t at 7pt}
\let\crnotice\mycrnotice%
\let\confname\myconfname%

\usepackage[english]{babel}
\usepackage[utf8]{inputenc}
\usepackage{amsmath}
\usepackage{graphicx}
\usepackage[colorinlistoftodos]{todonotes}
\usepackage{comment}
\usepackage{hyperref}

\title{Recommending Short-lived Dynamic Packages for \\ Golf Booking Services}

\date{\today}

\begin{document}

\newif\ifanonymous
\anonymousfalse

\newif\ifschedule
\schedulefalse

\newif\ifspace
\spacefalse

\newif\ifshort
\shorttrue

\ifanonymous
	\newcommand{\rakuten}{E-commerce Website Y }
	\newcommand{\gora}{[Anonymized] }
	\newcommand{\travel}{[Anonymized] }
\else
	\newcommand{\rakuten}{Rakuten Ichiba}
	\newcommand{\gora}{Rakuten GORA}
	\newcommand{\travel}{Rakuten Travel}

    \numberofauthors{2} 
    \author{
    Robin M. E. Swezey \qquad Young-joo Chung\\ 
             \affaddr{Rakuten Institute of Technology}\\
           \affaddr{1-14-1 Tamagawa, Setagaya-ku}\\
           \affaddr{Tokyo, Japan}\\
           \email{\{rswezey, yjchung\}@acm.org}
    }
\fi

\permission{Permission to make digital or hard copies of all or part of this work for personal or classroom use is granted without fee provided that copies are not made or distributed for profit or commercial advantage and that copies bear this notice and the full citation on the first page. Copyrights for components of this work owned by others than ACM must be honored. Abstracting with credit is permitted. To copy otherwise, or republish, to post on servers or to redistribute to lists, requires prior specific permission and/or a fee. Request permissions from Permissions@acm.org.}
\conferenceinfo{CIKM'15,}{October 19--23, 2015, Melbourne, Australia.} 
\copyrightetc{\copyright~2015 ACM. ISBN \the\acmcopyr}
\crdata{978-1-4503-3794-6/15/10\ ...\$15.00.\\
DOI: http://dx.doi.org/10.1145/2806416.2806608}

\clubpenalty=10000 
\widowpenalty = 10000

\maketitle

\begin{abstract}
We introduce an approach to recommending short-lived dynamic packages for golf booking services. Two challenges are addressed in this work. The first is the short life of the items, which puts the system in a state of a permanent cold start. The second is the uninformative nature of the package attributes, which makes clustering or figuring latent packages challenging. Although such settings are fairly pervasive, they have not been studied in traditional recommendation research, and there is thus a call for original approaches for recommender systems. In this paper, we introduce a hybrid method that leverages user analysis and its relation to the packages, as well as package pricing and environmental analysis, and traditional collaborative filtering. The proposed approach achieved appreciable improvement in precision compared with baselines.
\end{abstract}






\section{Introduction}
\label{sec:introduction}

Consider the booking or renting of a particular place on a reservation site. Specifically, we consider booking a round of golf through a golf booking service. Golf courses can be booked by users with a variety of options and prices: party size, a caddie to carry golf clubs, lunch, a guarantee of two-person \emph{pair} parties, a competition option for several parties, and a certain start time, among many other options. We would like to recommend to users not only courses they may like, but also priced packages of options that they may like on top of recommended parent courses.


Such options as described above are wide ranging. Extensive options are provided not only for golf courses on \gora, but hotels as well: rooms on \travel ~can be booked with many options ranging from breakfast/lunch/dinner to late checkout, and an outdoor bath. Additionally, coupons can be considered as short-lived dynamic packages that target regular E-commerce products. Car rental services also use such packages. In the present paper, we refer to these packages as \textit{short-lived dynamic booking packages}. Recommendations exist in the form of advertisement for this type of item in golf, but a preliminary survey\footnote{Survey conducted in 2013 on the following websites: \ifanonymous \emph{hidden for anonymity.} \else Alba (Japan), GDO (Japan), Golf-Jalan (Japan), GolfDigest (US), TeeOffTimes (UK).\fi} tells us that non-trivial recommendation systems seem not to exist. Considering the scale of the target industry given its user mass\footnote{On the order of 3 million users on \gora ~alone, approximately \ifanonymous 2.x\% of the population of country X.\else 2.5\% of the population of Japan.\fi} as well as its average order value\footnote{The AOV is approximately 90 USD per golf reservation\ifanonymous(USD used as anonymous currency)\fi.}, the recommendation of packages is essential for business success. However, to the best of our knowledge, such recommendations have not been studied in traditional research.

\begin{figure}[t]
\centerline{\includegraphics[width=0.5\textwidth]{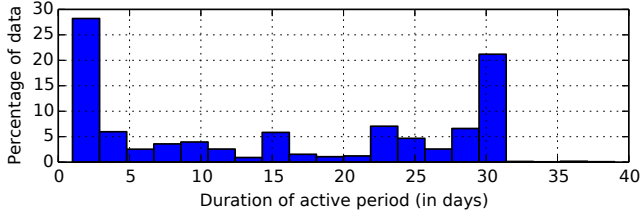}}
\caption{\label{fig:package_lifespans}Histogram of package lifespans in the period of June 2012 through May 2013.}
\end{figure}
\begin{figure}[t]
\centerline{\includegraphics[width=0.5\textwidth]{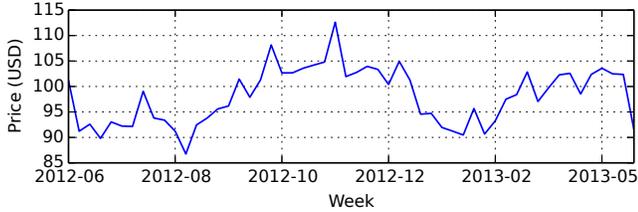}}
\caption{\label{fig:price_trends}Weekly price trends of packages.}
\end{figure}

There are \ifschedule three \else two \fi main challenges in recommending such packages. The first relates to their \emph{short-lived} aspect; on a B2B2C site, merchants (e.g., golf course owners or hotel owners) input the packages for their course and set different prices according to options, season, trends, and target customers.  We found that most such packages expire in a month after the start of their active period, including very short time-limited special offers (Figure~\ref{fig:package_lifespans}). Moreover, the price trends is what makes the packages \emph{dynamic} (Figure~\ref{fig:price_trends}). This puts the package recommendation system under a regime of a permanent cold start. Ratings, the objective variable most favored by classical collaborative filtering approaches for atomic items, are not available. Co-counting using purchasing/browsing history is also very limited, as customers book an average of 4.5 courses per year, i.e., one package every 2.7 months. This means that they do not book packages fast enough on average for traditional models to be learned and used in the short term.

In recommendation, the straightforward approach under a cold start regime or working on long-tailed items is to turn to content-based methods of information retrieval. This gives rise to the second challenge: uninformative data. Package contents comprise flags and categorical variables for various options, but analysis of the items alone based on their content results in poor clusters or latent packages, because options have different importance to the package value from the point of view of the user. Thus, direct application of clustering using similarities such as the Jaccard index, where every attribute is weighed the same, performs poorly.




In addressing the challenges mentioned, we leverage reservation histories enriched with package and course data to assess user behavior, and conduct an analysis of package pricing. This allows us to construct a similarity score that performs well. Our approach is threefold, in that we:

1. extract user behavioral characteristics,

2. conduct collaborative filtering on parent items, and

3. perform content-based information retrieval using user preferences and the package price.


\section{Recommending Golf Packages}

\subsection{User-weighed package similarity}

\label{sec:preliminary_analysis}
\label{sec:clustering}

\begin{figure}[t]
\centerline{\includegraphics[width=0.5\textwidth]{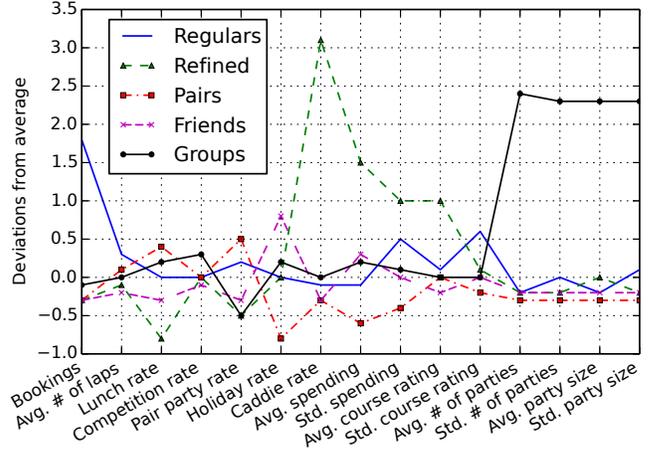}}
\caption{\label{fig:clusters}Cluster centroids of users having booked at least two courses. $Pairs$ and $friends$ each comprise 35\% of users, while the others each comprise 10\%.}
\end{figure}


Plan data themselves are uninformative for similarity (as shown in Section ~\ref{sec:experiments} by Jaccard's poor precision). We analyze packages through users, enriching their reservation history with package and course data. The data were aggregated to build user behavior vectors, which we Z-transformed and clustered using Euclidean k-means. Figure~\ref{fig:clusters} shows vastly different behaviors with regard to package options and price (e.g., spending deviation growing with the spending average), and a need to develop adequate similarity metrics.
We define the user-weighed option similarity score for a package $p$ with respect to a user $u$ as

\iffalse

\begin{equation}
\tilde{S}_{opt}(p|p^{(ref)},u) = \sum_{k \in O} w_k P(p_k|p^{(ref)},u)
\end{equation} \\
where $w_k$ is a weight, $k$ belongs to the subset $O$ of vector indices of $p$ denoting categorical option attributes (which are binary or dummy-coded). P is defined as:\\
\begin{equation}
P(p_k|p^{(ref)},u) = \frac {1}{1+e^{-(\beta_{0,S_u} + \beta_{1,S_u}^T p^{(ref)})}},
\end{equation} \\

\else
\[
\tilde{S}_{opt}(p|u) = \sum_{k \in O} P(p_k|u)
\]
\noindent where $k$ belongs to the subset $O$ of vector indices of $p$ denoting flags or dummy-coded categorical attributes, and P is a logistic factor such as \\
\[
P(p_k|u) = \frac {1}{1+e^{-(\beta_{0,S_u} + \beta_{1,S_u}^T u)}},
\]
\fi

\noindent where $\beta_{0,S_u}$ and $\beta_{1,S_u}$ are respectively the intercept and coefficient vector for cluster $S_u$ to which $u$ belongs after clustering. This probability corresponds to user $u$ choosing the option $p_k$ for the next booking. $P(P_k|U)$ is learned by leaving the last package $p^{(last)}$ booked by a user and using its $p_{k \in O}^{(last)}$ as dependent variables and the user vector as independent variable for learning.
We show example logistic weights for three output probabilities in Table ~\ref{tab:logistic_models}, each predicting the future occurrence of an option\footnote{We used Weka version 5.3.001 for computing logistic regression (http://www.cs.waikato.ac.nz/ml/weka/).}. 

\begin{table}[tbp]
\caption{\label{tab:logistic_models}Weights of logistic models predicting occurrence of options in the next booking. Only weights that are over $10^{-1}$  for a response (in bold) are shown.}
\begin{center}
\begin{tabular}{lrrr}
\cline{2-4}
 & \multicolumn{3}{c}{Options}\\ \hline
User attribute & \multicolumn{1}{l}{Caddie} & \multicolumn{1}{l}{Holiday} & \multicolumn{1}{l}{Lunch} \\ \hline
Lunch rate & \textbf{-0.315} & \textbf{-0.1829} & \textbf{1.4113} \\ 
Competition rate & 0.0351 & -0.0166 & \textbf{0.1656} \\ 
Holiday rate & \textbf{-0.351} & \textbf{1.6612} & \textbf{-0.1516} \\ 
Caddie rate & \textbf{3.0924} & \textbf{-0.6099} & \textbf{-0.1776} \\ 
Avg. spending & \textbf{0.201} & 0.092 & \textbf{-0.103} \\ 
Avg. course rating & \textbf{0.3673} & \textbf{-0.3004} & \textbf{0.2138} \\ 
Std. course rating & \textbf{0.295} & \textbf{-0.3063} & 0.0637 \\ 
Avg. \# of parties & \textbf{-0.9268} & \textbf{-0.1061} & \textbf{-0.1751} \\ 
Std. \# of parties & \textbf{0.1424} & 0.0083 & \textbf{0.117} \\ 
Avg. party size & \textbf{0.2164} & 0.0064 & 0.0667 \\ 
Intercept & \textbf{-5.6731} & \textbf{-0.6552} & 0.0487 \\ \hline
\end{tabular}
\end{center}
\end{table}

\subsection{Reference course and package}
\label{sec:reference_selection}

The reference package is used mainly to compute the price similarity described in Section~\ref{sec:price_similarity}. It is also necessary to perform the subsequent experiments of this paper for the Jaccard baseline (Section~\ref{sec:experiments}). We first extract the reference course that was played the most in the season closest to the target season, using a simple scoring function. For example, if we would like to recommend packages in June, a course played twice around June scores higher than one played twice in December. This is based on the assumption that a user likes a course if he has booked it several times, and the observation that users have affinities to courses that are seasonal. Once the reference course is selected, we simply select the last package booked as the reference package.

\subsection{Filtering of parent items}
\label{sec:course_filtering}

Collaborative filtering should be leveraged wherever possible, even if impractical for the granularity of packages. In our case, there are \ifanonymous 2,000 courses\footnote{Exact number hidden for anonymity.} \else 1,951 courses \fi in our system, each generating packages that we want to recommend on top of them to users.
This gives us a parent course item/item co-occurrence matrix for courses of rank \ifanonymous 2,000\else 1,951\fi. Because we have at least 100,000 active users in the period to populate the co-occurrence matrix, the matrix is very dense, which works well with collaborative filtering. As the course recommender already performs well on \gora, we choose to use it in a collaborative filtering step to filter courses of interest to the user.

\subsection{Price similarity}
\label{sec:price_similarity}

To improve the scoring function, we develop a similarity component based on package price, which should be leveraged because our analyses reveal two important patterns:

1. 90\% of users do not deviate by more than 30\% from their average spending  (the remaining 10\% belong to the cluster of \emph{refined} users in Figure~\ref{fig:clusters}), and

2. the price itself contains enough information about the package to make it a potent similarity measure.

\begin{figure}[t]
\centering
\includegraphics[width=0.5\textwidth]{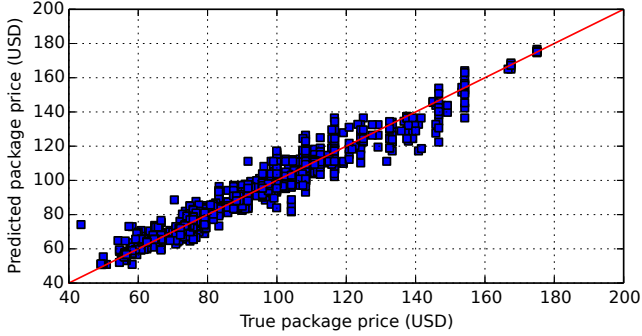}
\caption{\label{fig:price_regression}True prices vs. predictions for a course.}
\end{figure}

\begin{table}[tbp]
\caption{\label{tab:regression_features}Sets of features for the linear model of price. The final set is $\{m\} \times \{d\} \times A \times P$.}
\begin{center}
\begin{tabular}{ll}
\hline
\multicolumn{1}{l}{Set} & \multicolumn{1}{l}{Features} \\ \hline
Temporal & Month of year (m), day of week (d) \\
Attributes (A) & Lunch, caddie, competition, \\
           & pair party, min. party size, \\
           & min. nb. of parties, nb. of laps \\
Promotional (P) & Promotion type, shortness of package \\
\hline
\end{tabular}
\end{center}
\end{table}

To demonstrate point (2), we run a regression on a course in the data that has generated many packages. Using the Cartesian feature set of Table~\ref{tab:regression_features}, we build a linear model that gives prices of packages that are fairly close to the truth (see Figure~\ref{fig:price_regression}). Note that the plot in Figure~\ref{fig:price_regression} is heteroscedastic, which shows that pricing becomes loose at the high end, as consistently does spending (Section~\ref{sec:clustering}).
We define price similarity score for a package $p$ with respect to a user $u$ and a reference plan $p^{(ref)}$ selected from his/her history as
\[
\tilde{S}_{price}(p|p^{(ref)},u) = \frac{1}{1 + \frac{r_{t_p,t_{p^{(ref)}}}}{\omega + \sigma_{price}^{(u)}} \| price_p - price_{p^{(ref)}} \| },
\]
where $r_{t_p,t_{p^{(ref)}}}$ is the ratio of seasonal averages between $p$ and $p^{(ref)}$ compensating for seasonal trends, $\omega$ is a currency scaling factor, and $\sigma_{price}^{(u)}$ is the user's spending deviation.

\subsection{Final score}
\label{sec:final_score}

The final score is defined as
\[\tilde{S}(p|p^{(ref)},u) = ( w_p \tilde{S}_{price} + w_o \tilde{S}_{opt} + w_c \tilde{S}_c ) [ p|p^{(ref)},u ],
\]
where $\tilde{S}_c$ is the parent course score after filtering (Section~\ref{sec:course_filtering}), and $w_p$, $w_o$ and $w_c$ are weights whose optimal values can be found through hill-climbing with respect to EMP@n.

\section{Experiments}

This section details the offline and online evaluation. For offline evaluation, we compared the precision of our approach with that of the basic similarity method, and tested three different options. We then launched an e-mail campaign for online evaluation. None of our data contains personal information such as names or addresses. 

\subsection{Offline evaluation}
\label{sec:experiments}

We tested our proposed methods on golf booking data collected from June 2012 to May 2013. The number of unique users in this period is 521,442 and the total number of bookings is 2,499,678. We used the booking history from June 2012 to May 2013 to generate the package recommendations. We then checked what packages users actually booked from 1 June 2013 to 15 June 2013. We call this evaluation index the \textit{expected minimum precision} (EMP), because the number of booked packages would increase if users interact with our recommendation results. This setting has been widely used in the evaluation of recommender systems; e.g.,~\cite{zhu2014bundle}. The EMP is defined as
\[
P_n = \sum_{u \in U} \frac{| recommendations_{u,n} \cap truth_u |}{| truth_u |}.
\]

We test four settings in this experiment:

1. the Jaccard score computed as
\[
S_{J}(p,p^{(ref)}) = \frac{|p \cap p^{(ref)}|}{|p \cup p^{(ref)}|}
\]
in place of $S_{price}$ and $S_{opt}$, where the $p$ values are here used to denote the attribute sets,

2. the user-weighed option similarity $S_{opt}$ without $S_{price}$,
 
3. the final score incorporating $S_{opt}$ and $S_{price}$ without $r_{t_p,t_{p^{(ref)}}}$ for price adjustment, and

4. the final score incorporating $S_{opt}$ and $S_{price}$ with $r_{t_p,t_{p^{(ref)}}}$ for price adjustment.
\\
Each of these settings incorporate the same reference course and package selection step (Section~\ref{sec:reference_selection}) and course filtering step (Section~\ref{sec:course_filtering}).

\begin{figure}[t]
\centering
\includegraphics[width=0.5\textwidth]{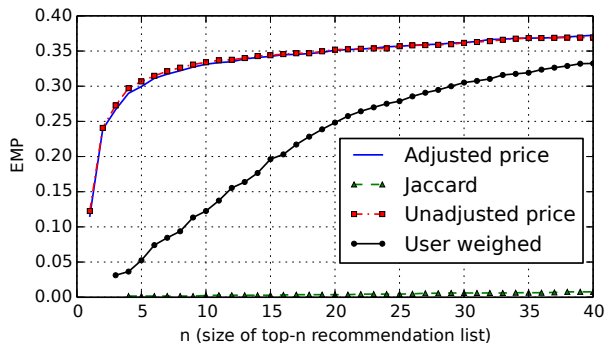}
\caption{\label{fig:EMP}EMP@n curves for each tested method.}
\end{figure}

Figure~\ref{fig:EMP} shows that the scores incorporating price similarity perform best, independently of price trend adjustment. User-weighed similarity performs reasonably well but not nearly as well as when price is incorporated. Finally, the Jaccard baseline shows very poor performance. For a top-five returned list, the EMP of the proposed method is 25\% more efficient than that of user-weighed similarity, and 30\% more than that of the Jaccard baseline.

\subsection{Online Evaluation}
We performed online evaluation by conducting a personalized e-mail campaign. We sent e-mails that contained six recommended packages on one day in December 2014 and evaluated the click-to-open-rate (CTOR) and conversion-to-open-rate (CVR). Here, the CVR refers to the event that a customer clicked on the recommended package and made a reservation.  We compared the performance to that of previous and following e-mail campaigns. E-mails in these two campaigns contained approximately 100 packages selected by specialists based on their contents and target demographics. Our approach achieved the highest CTOR and CVR with the maximum improvement in the CTOR was 200\%.

\section{Related Work}
\label{sec:related_work}


We briefly describe previous research on dynamic items and a cold start. 
Schein et al.~\cite{Schein:2002:MMC:564376.564421} raised the cold-start problem in recommendation. To overcome the lack of user preference information that is essential for collaborative filtering, they proposed a probabilistic model that combines content and collaborative filtering. Chu and Park~\cite{Chu:2009:PRD:1526709.1526802} combined user and item profiles in a dynamic bilinear model for time-aware recommendation for the Today module on the Yahoo! front page. Matrix factorization techniques that solve a cold-start problem were also proposed in several works~\cite{Koenigstein:2011:YMR:2043932.2043964,Saveski:2014:ICR:2645710.2645751}.

The crucial difference between short-lived dynamic items and the items in the research body on permanent cold-start regimes is their short life-span and non-retrievable aspects. To be more specific, \textit{short-lived} items  will expire within a month after the start of their active period, as is shown by our observations (Figure~\ref{fig:package_lifespans}). Such items cannot be retrieved (i.e., booked, searched for, browsed, or recommended) by/to users once they expire. They also have generally poor statistical value in a user/item matrix because of their capped counts, especially when this matrix does not embody a notion of time and relevance to the present.

On the other hand, Zhu et al.~\cite{zhu2014bundle} proposed the bundle recommendation problem in e-commerce. According to the observation that users usually buy more than one item on e-commerce sites and that displaying related items together improves conversion, they proposed a recommender system that maximizes the reward function (conversion rate and revenue/profit of the bundle). This work is different from ours in that it focused on creating bundles, whereas our work focuses on recommending packages that are created by merchants using different values of limited attributes such as lunch, and competition.

\section{Conclusion and Future Works}
\label{sec:conclusion}
\label{sec:future_work}

In this work, we identified a pervasive subset of items that ought to be the subject of a recommendation research: short-lived dynamic booking packages. We showed that the problem is unique in terms of the short lifespan of items and the uninformative nature of the data, which calls for an original recommendation approach. We performed an experiment over a subset of users that resulted in appreciable improvements in EMP when leveraging user analysis and price analysis of the package to define adequate scoring functions. We also performed an actual A/B test for a mail recommendation and found that the click-to-open rate was twice that achieved with human selection of packages.
In further work, we would like to refine the metrics and address a third challenge, namely that booking packages are designed to \emph{book} a parent item, which inherently adds a notion of \emph{schedule} to the problem constraints and makes recommendation challenging when the user schedule is unknown.

\ifanonymous
\else
\section*{Acknowledgements}
We would like to thank Satoko Marumoto, Takahiro Kuroda, Yusuke Sasamori, Ryo Yoneda, Yoshiro Matsuda, Yu Hirate, and all contributors to this research for their support.
\fi

\bibliographystyle{acm}
{\small
\bibliography{sigproc.bib}
}

\end{document}